\documentclass[aps,prl,reprint,twocolumn,showkeys,showpacs,preprintnumbers,amsmath,amssymb,floatfix]{revtex4-1}

\usepackage{graphicx}
\usepackage{dcolumn}
\usepackage{bm}
\usepackage{dsfont}
\usepackage{textcomp}

\begin{document}
\title{Real Time Electrical Detection of Coherent Spin Oscillations}

%Electrically Detected Pulsed ENDOR in Phosphorus-Doped Silicon
\author{Felix Hoehne}
\email[corresponding author, email: ]{hoehne@wsi.tum.de} 

\author{Christian Huck}
\affiliation{Walter Schottky Institut, Technische Universit\"{a}t
M\"{u}nchen, Am Coulombwall 4, 85748 Garching, Germany}

\author{Hans Huebl}
\affiliation{Walther-Mei\ss ner-Institut, Bayerische Akademie der Wissenschaften, Walther-Mei\ss ner-Str.\,8, 85748 Garching, Germany}
\author{Martin S.~Brandt}
\affiliation{Walter Schottky Institut, Technische Universit\"{a}t
M\"{u}nchen, Am Coulombwall 4, 85748 Garching, Germany}
\begin{abstract}
We demonstrate that the bandwidth of pulsed electrically detected magnetic resonance can be increased to at least 80~MHz using a radio frequency-reflectometry detection scheme. Using this technique, we measure coherent spin oscillations in real time during a resonant microwave pulse. We find that the observed signal is in quantitative agreement with simulations based on rate equations modeling the recombination dynamics of the spin system under study. The increased bandwidth opens the way to electrically study faster spin-dependent recombination processes, e.g., in direct semiconductors which so far have almost exclusively been studied by optically detected magnetic resonance.
\end{abstract}


\maketitle

%I
Recombination processes are ubiquitous in bipolar semiconductor devices such as inorganic or organic light emitting diodes and solar cells. Particularly valuable information can be obtained when a recombination process is spin-dependent since this allows for the spectroscopic identification of the participating charge carriers, recombination centers or charge transfer complexes via their spin signatures~\cite{Dersch83,Chen1991,Carlos1995,Dyakonov1996,Stutzmann2000} by using methods such as optically or electrically detected magnetic resonance (ODMR and EDMR, resp.)~\cite{Greenham1996,Spaeth03}. In addition, by means of coherent spin manipulation and pulsed optical excitation of charge carriers, highly relevant information on charge carrier dynamics can be obtained, allowing to determine, e.g., trapping and recombination times~\cite{Morley2008,Hoehne_Timeconstants_2013}. To this end, complex sequences consisting of microwave (mw) pulses for electron spin manipulation, radiofrequency (rf) pulses for nuclear spin manipulation and light pulses for carrier excitation have been developed~\cite{Childress06NVQC,Dreher2012}.
However, in the case of pulsed EDMR the finite bandwidth of conventional preamplifier-based current measurement setups limits the time resolution to some microseconds. For the observation of phenomena faster than that like coherent spin oscillations or fast recombination processes one therefore resorts to an indirect detection technique which allows to reconstruct the state of the different spin ensembles relevant for the recombination by measuring the spin-dependent part of the current transient after the pulse sequence~\cite{Boehme03EDMR}. If, e.g., the coherent driving of a particular spin ensemble in a Rabi oscillation experiment is to be monitored, this requires the time-consuming measurement of a separate transient for each driving pulse length followed by a reconstruction of the Rabi oscillation from an analysis of these transients~\cite{Stegner06,machida2003}. Moreover, this method is only applicable to spin systems where at least one of the spin-dependent time constants is sufficiently long to be detected with the available measurement bandwidth. For continuous wave (cw) EDMR, it has been demonstrated~\cite{Huebl09} that the detection bandwidth can be increased by more than one order of magnitude employing an rf-reflectometry-based detection scheme~\cite{Schoelkopf98,Angus08SET} which simultaneously improves the signal-to-noise ratio by avoiding low-frequency noise. Here, we combine this detection scheme with pulsed spin manipulation and use it to observe coherent spin oscillations in real time during the mw pulse, in contrast to the reconstruction from the photocurrent transient after the pulse. Furthermore, with the help of a quantitative model we show that the signal intensity of real time pulsed EDMR and its time dependence are in very good agreement with the results of the conventional pulsed EDMR, demonstrating that we now have an additional highly versatile method at our hands to characterize fast charge and spin dynamics in semiconductors down to nanosecond time scales.

\begin{figure}[!t]
\begin{centering}
\includegraphics[width=0.8\columnwidth]{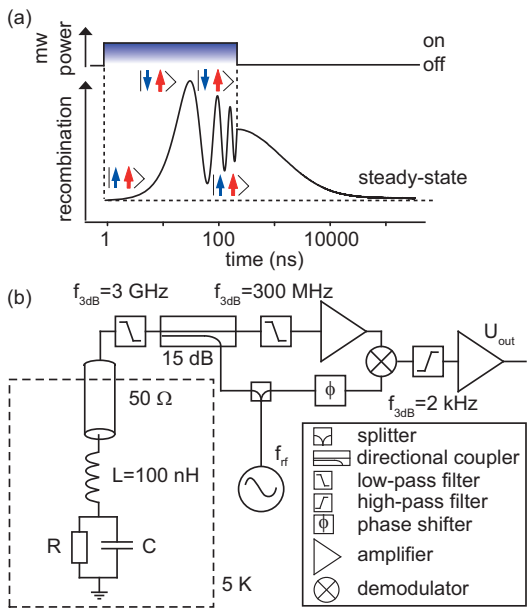}
\par\end{centering}
\caption{\label{fig:Figure1} 
(a) Basic concept of a pulsed EDMR measurement. The symmetry of the spin pair determines the recombination rate with parallel spin pairs recombining much slower than antiparallel spin pairs due to the Pauli principle. While conventional pulsed EDMR can only assess the state of the spin pair at the end of the mw pulse by measuring the photocurrent transient, rf-EDMR can directly monitor the coherent spin oscillations during the mw pulse.
%The state of the spin pair (blue and red arrow) after a resonant mw pulse can be accessed by box-car integrating the current transient following the mw pulse, the amplitude of which is proportional to the number of antiparallel spin pairs at the end of the mw pulse. 
%Before the microwave pulse, the spin pairs are in the long-lived antiparallel configuration. A microwave pulse resonant with one of the two spins induces coherent spin oscillations which change the symmetry of the spin pair between antiparallel and parallel which in turn leads to a variation of the recombination rate. However, the bandwidth of conventional EDMR current measurement setups is not sufficient to resolve these oscillations which take place on a timescale of less than a microsecond. Nevertheless, the state of the spin pair after the microwave pulse can be accessed by box-car integrating the current transient following the microwave pulse (gray shaded area), the amplitude of which is proportional to the number of antiparallel spin pairs at the end of the mw pulse.
(b) Schematic of the rf-EDMR LCR tank circuit and homodyne detection setup. 
%The sample is connected to the inner and outer conductors of a 50~$\Omega$ coplanar waveguide (CPW) with a chip inductance of $L$=100~nH placed between the sample and the inner conductor. The sample resistance $R$, its stray capacitance $C$, and the inductance $L$ form a resonant LCR circuit. The reflected power is measured with a standard homodyne detection setup with a 15~dB directional coupler, low-pass filters, amplifiers, and a demodulator for downconversion.
%The demodulated signal $U_\mathrm{out}$ is further filtered, amplified and recorded with a lock-in amplifier (cw measurements) or a digital sampling card (pulsed measurements).
}
\end{figure}

Before describing the pulsed rf-reflectometry EDMR (rf-EDMR) measurements, we briefly review the principle of pulsed EDMR measurements in a little more detail~\cite{Boehme03EDMR,Stegner06}. Most EDMR signals can be described in terms of weakly coupled spin pairs, where the recombination rate between two paramagnetic localized states depends on the relative orientation of the two spins [red and blue arrow in Fig.~\ref{fig:Figure1}(a)]~\cite{Kaplan78Spindep}. Spin pairs with an antiparallel orientation of the two spins recombine rapidly, while parallel spin pairs are stable on a much longer timescale. Therefore, under above-bandgap illumination a steady-state develops with almost all spin pairs in the parallel state. Resonant excitation of one of the two spins by mw irradiation increases the number of antiparallel spin pairs and consequently also the recombination rate which results in a resonant decrease of the photoconductivity. In the most simple pulsed EDMR experiment illustrated in Fig.~\ref{fig:Figure1}(a), a resonant mw pulse causes one of the two spins (blue arrow) to coherently oscillate between its eigenstates. This changes the symmetry of the spin pair resulting in an oscillation of the overall recombination rate. The frequency of this oscillation (tens of MHz) is chosen much faster than the typical decoherence rates~\cite{Huebl08Echo} and, therefore, in many cases larger than the bandwidth of most EDMR current detection setups (usually below 1~MHz) preventing the direct observation of these oscillations. However, the amplitude of the current transient after the mw pulse is proportional to the number of antiparallel spin pairs at the end of the pulse, so that the state of the spin pair can be determined by measuring the current transient~\cite{Boehme03EDMR}. In the following, we demonstrate that the limitations of this indirect detection scheme can be overcome by rf-reflectometry allowing to detect the coherent spin oscillations during the mw pulse. 

%Figure 1
%Figure 1

The samples used in this work were grown by chemical vapor deposition and consist of a nominally 
$22$~nm thick Si layer with a P concentration of $\sim3\cdot10^{16}$~cm$^{-3}$ on a 
$2.5$~\textmu m thick, undoped Si buffer grown on a (100)-oriented silicon-on-insulator 
substrate. The doped epilayer leads to a dominant $^{31}$P-P$_\mathrm{b0}$ recombination~\cite{Hoehne10}, where the P$_\mathrm{b0}$ spin partners are defect states at the interface of the doped epilayer and the natural oxide formed on top~\cite{Stesmans98dbHF}. All experiments are performed at 5~K under illumination with red light of an LED (photon energy $h\nu=1.95$~eV)
in a dielectric mw resonator for pulsed EPR at X-band frequencies. Interdigit Cr/Au electrical contacts with a periodicity of 10~\textmu m are evaporated on an area of 2x2~mm$^2$. 

\begin{figure}[!t]
\begin{centering}
\includegraphics[width=\columnwidth]{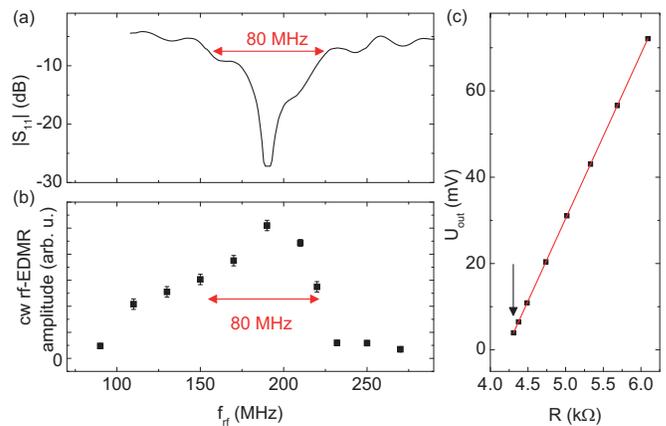}
\par\end{centering}
\caption{\label{fig:Figure2} 
(a) Absolute value of the reflection coefficient $\left|S_{11}\right|$ as a function of the rf frequency $f_\mathrm{rf}$. From the data we extract a resonance frequency of $f_0$=190~MHz and a bandwidth of $\sim$80~MHz (full width at half maximum) for the LCR resonator as indicated by the red arrow. 
(b) Amplitude of the high-field $^{31}$P hyperfine peak measured by cw rf-EDMR as a function of $f_\text{rf}$. The bandwidth of the LCR resonator determined in (a) is shown by the red arrow for comparison.
(c) Calibration of the rf-reflectometry setup. The output voltage $U_\mathrm{out}$ exhibits a linear dependence (red line) on $R$ around the working point indicated by the arrow.
}
\end{figure}

For rf-reflectometry, a chip inductance of $L$=100~nH is placed between the sample and a 50~$\Omega$ coplanar waveguide (CPW), which connects the sample to the room-temperature electronics via a 50~$\Omega$ coaxial cable [Fig.~\ref{fig:Figure1}(b)]. The sample resistance $R$, its stray capacitance $C$ and the inductance $L$ form a resonant LCR tank circuit with a resonance frequency of $f_0\approx1/\sqrt{LC}$ whose impedance can be matched to 50~$\Omega$  by varying $R$ via the illumination intensity. Measuring the reflected rf power as a function of the radio frequency $f_\mathrm{rf}$ using a vector network analyzer, we find a resonance frequency of $f_0$=190~MHz and a bandwidth (FWHM) of $\sim$80~MHz [Fig.~\ref{fig:Figure2}(a)]. Note, that we have designed the frequency of the LCR tank circuit to avoid frequencies corresponding to nuclear magnetic resonance transitions in the spin system studied. For rf-EDMR measurements, we use the rf-reflectometry homodyne detection setup shown in Fig.~\ref{fig:Figure1}(b). It is calibrated
for resistance measurements by simultaneously measuring the output voltage of the demodulator $U_\mathrm{out}$ at $f_\mathrm{rf}$=$f_0$=190~MHz and the DC sample resistance $R$ as a function of the illumination intensity. From this, we obtain a relation between $U_\mathrm{out}$ and $R$ as shown in Fig.~\ref{fig:Figure2}(c) revealing a linear dependence around the working point at $R$=4250~$\Omega$ indicated by the arrow.
The shape of the resonant dip [Fig.~\ref{fig:Figure2}(a)] deviates from the expected Lorentzian shape mostly likely due to spurious reflections at the transitions between the coaxial cable and the CPW and between the CPW and the sample. From the resonance frequency and the value of the inductance, we calculate a capacitance of $C=1/L(2\pi f_0)^2=7$~pF in good agreement with the estimated capacitance of the interdigit contact structure of $\sim$14~pF~\cite{Endres91}.

\begin{figure}[!t]
\begin{centering}
\includegraphics[width=0.8\columnwidth]{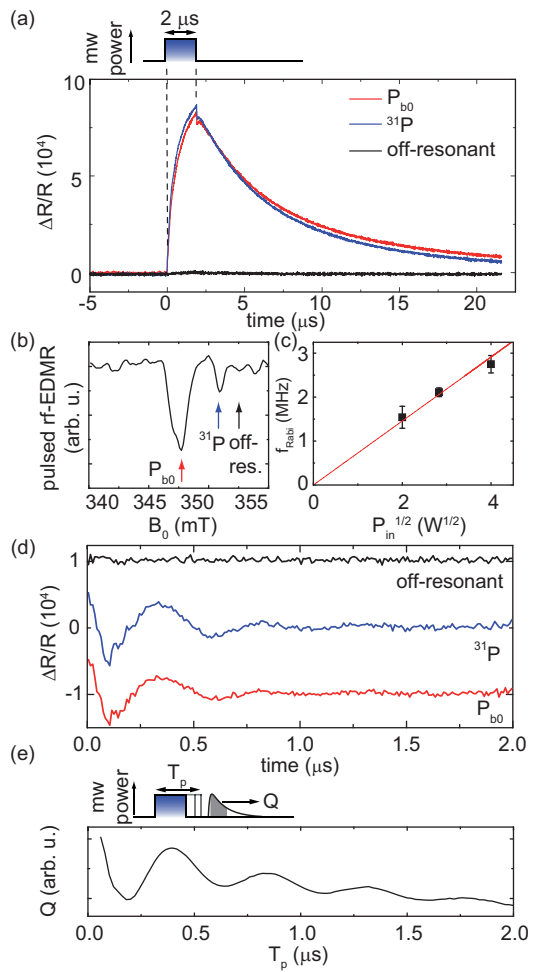}
\par\end{centering}
\caption{\label{fig:Figure3} 
(a) Relative change of the sample resistance for different magnetic fields chosen such that the mw pulse resonantly excites the $^{31}$P spins (blue), the $P_\text{b0}$ spins (red) or none of the spins (black). To improve the signal-to-noise ratio, a lock-in detection scheme~\cite{Hoehne2012} is implemented employing square-wave frequency modulation with a frequency of 500~Hz.
(b) Pulsed rf-EDMR spectrum recorded by box-car integration of the $\Delta R/R$ transient after a 9.739~GHz mw pulse as a function of the magnetic field $B_0$. The spectral positions of the mw pulses used in (a) are indicated by the color-coded arrows.
(c) Frequency of the oscillations $f_\text{Rabi}$ depicted in (d) as a function of the square-root of the mw power $P_\mathrm{in}$ showing a linear dependence (red line). 
(d) First two microseconds of the transients from (a) after subtraction of a second order polynomial background revealing oscillations on the two resonant traces.
(e) Coherent spin oscillations measured by reconstructing the spin state after the mw pulse of length $T_\mathrm{p}$~\cite{Stegner06}. To this end, the preamplifier-detected current transient after each pulse is integrated resulting in a charge $Q$ proportional to the number of antiparallel spin pairs at the end of the pulse.
}
\end{figure}

In a next step, we use this measurement scheme to detect the change of $R$ induced by the resonant excitation of $^{31}$P spin transitions in cw rf-EDMR. For this purpose, the sample is continuously irradiated with microwaves with the frequency of 9.739~GHz chosen such that the spectrally isolated high-field $^{31}$P hyperfine-split electron spin transitions is resonantly excited at a magnetic field of $B_0$=350.6~mT [blue arrow in the spectrum in Fig.~\ref{fig:Figure3}(b)]. 
%The resonant change of $R_\text{sample}$ is observed in cw rf-reflectometry EDMR~\cite{Huebl09} by measuring the reflected rf power as a function of the magnetic field for different radio frequencies $f_\text{rf}$. For these measurements, we employ lock-in detection of the downconverted rf signal in combination with magnetic field modulation at a frequency of 1.234~kHz to improve the signal-to-noise ratio. 
The amplitude of the $^{31}$P signal shown in Fig.~\ref{fig:Figure2}(b) as a function of $f_\text{rf}$ is maximal for $f_\text{rf}$=$f_0$=190~MHz and decreases to almost zero for $f_\text{rf}$\textgreater 250~MHz or $f_\text{rf}$\textless 100~MHz. These results directly reflect the frequency-dependent sensitivity of the rf-reflectometry setup which is maximal when $f_\text{rf}$ matches the resonance frequency of the LCR resonator and close to zero for $f_\text{rf}$ far away from the resonance~\cite{Schoelkopf98,Angus08SET}. The frequency range over which an EDMR signal is observed [red arrow in Fig.~\ref{fig:Figure2}(b)] agrees well with the bandwidth of 80~MHz determined in Fig.~\ref{fig:Figure2}(a) confirming that the rf-reflectometry indeed should allow EDMR measurements with a time resolution of tens of nanoseconds.

%We further calibrate the output of the rf reflectometry setup by measuring the output voltage of the IQ demodulator $U_\mathrm{out}$ at $f_\mathrm{rf}$=190 MHz as a function of the illumination intensity, which in turn is translated into a change of resistance by measuring the DC sample resistance $R_\text{sample}$ also as a function of the illumination intensity. From this, we obtain a relation between $U_\mathrm{out}$ and $R_\text{sample}$ as shown in Fig.~\ref{fig:Figure2}(c) revealing an almost linear dependence around the working point at $R_\text{sample}$=4250~$\Omega$ indicated by the arrow.  

%Figure 1

In the following, we use the large detection bandwidth of rf-EDMR to observe coherent spin oscillations during the mw excitation pulse as summarized in Fig.~\ref{fig:Figure3}. To this end, we irradiate the sample with a 2~\textmu s long mw pulse at the fixed frequency of 9.739~GHz and simultaneously measure the time dependence of $R$ during and after the mw pulse using the calibration of Fig.~\ref{fig:Figure2}(c). 
%To improve the signal-to-noise ratio, a lock-in detection scheme~\cite{Hoehne2012} is implemented by repeating the mw pulse with a shot repetition time of 2~ms while the mw frequency is switched between two values differing by 40~MHz, much more than the linewidths of 12~MHz and 20~MHz for the $^{31}$P and P$_\text{b0}$ peaks, respectively. This effectively results in a square-wave modulation of the signal with a frequency of 500~Hz, which is detected by subtracting the signals from two consecutive cycles. In addition, the non-resonant part of the current transient caused by the strong mw pulses is removed by recording transients at two additional values of the magnetic field, where no resonant signal is observed. For each of the transients shown in Fig.~\ref{fig:Figure3}(a), the linearly interpolated off-resonant transient is subtracted. 
The results in Fig.~\ref{fig:Figure3}(a) show the relative change of the sample resistance for three different values of $B_0$. Two of the values are chosen such that the mw pulse resonantly excites the $^{31}$P and P$_\mathrm{b0}$ transitions (blue and red trace), while the third value is chosen off-resonant for comparison (black trace). The corresponding spectral positions are indicated by the according color-coded arrows in the pulsed rf-EDMR spectrum shown in Fig.~\ref{fig:Figure3}(b). The resistance first increases during the mw pulse and decreases after the pulse with a time constant of $\sim$5~\textmu s for the two resonant transients, while no variation is observed in the off-resonant transient [Fig.~\ref{fig:Figure3}(a)]. The maximum value of $\Delta R/R \approx 7\cdot 10^{-4}$ is comparable to the maximum change of $\Delta R/R\approx 10^{-3}$ observed in conventionally detected pulsed EDMR experiments on this sample. 
During the mw pulse, a weak oscillation is present on the two resonant traces, which is revealed after subtraction of a second order polynomial background as shown in Fig.~\ref{fig:Figure3}(d). Oscillations with a period of 500~ns are present in both resonant traces, while they are not observed for the off-resonant trace. We attribute these oscillations to the changes in the recombination rate caused by coherent spin oscillations during the mw pulse observed in real time [Fig.~\ref{fig:Figure1}(a)]. For comparison, coherent oscillations measured by conventionally detected pulsed EDMR~\cite{Stegner06} are shown in Fig.~\ref{fig:Figure3}(e), exhibiting the same oscillation frequency as those measured by rf-reflectometry. Our interpretation is further confirmed by the linear dependence of the oscillation frequency $f_\text{Rabi}$ of the pulsed rf-EDMR on the square-root of the mw power [Fig.~\ref{fig:Figure3}(c)].

The oscillation amplitude of $\sim 5\cdot 10^{-5}$ is much smaller than the overall resonant resistance change of $\sim 7\cdot 10^{-4}$. The small amplitude of the oscillation results from two conflicting conditions which have to be met in order to observe the recombination process. On the one hand, the recombination has to be sufficiently fast compared to the oscillation period of the spin, so that the change of the spin pair state is reflected instantaneously in the photocurrent. On the other hand, however, the spin pair is destroyed by the recombination process leading to a rapid decay of the oscillation for recombination processes much faster than the oscillation period.

%---------------------------figure-------------------------------
\begin{figure}[!t]
\begin{centering}
\includegraphics[width=0.8\columnwidth]{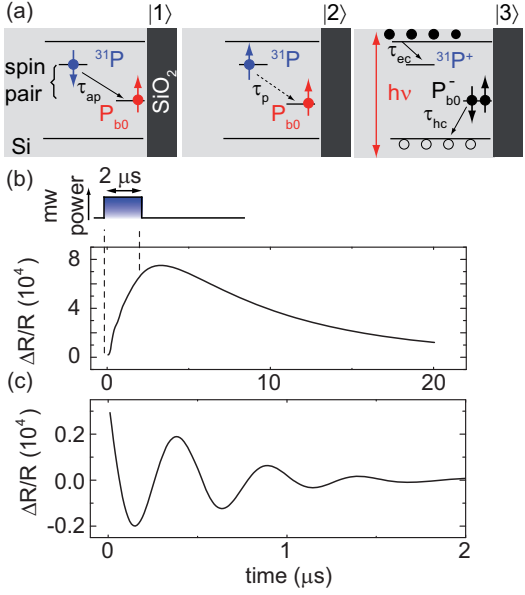}
\par\end{centering}
\caption{\label{fig:Figure4} 
(a) Definition of the time constants of the $^{31}$P-P$_\mathrm{b0}$ recombination process. 
(b) Simulation of the relative resistance change during and after a 2~\textmu s long mw pulse.
(c) First 2~\textmu s of the data shown in panel (a) after subtraction of a second order polynomial background.
}
\end{figure}
%-----------------------------figure------------------------------------
To gain quantitative insight into this, we calculate the change of the photoconductivity by modeling the dynamics of the spin pair based on a set of rate equations. We hereby extend the approach of Ref.~\cite{Hoehne_Timeconstants_2013} by explicitly including the dynamics of the spins during the mw pulse.
We discuss the dynamics of the spin pair using three states, namely the parallel state of the spin pair $\left|1\right\rangle$, the antiparallel state $\left|2\right\rangle$, and the ionized $^{31}$P$^+$-P$_\mathrm{b0}^-$ state $\left|3\right\rangle$ as sketched in Fig.~\ref{fig:Figure4}(a). The time evolution of the corresponding density matrix $\rho(t)$ is calculated by a master equation
\begin{equation}
\label{eq:Liouville_I}
\frac{\mathrm{d}\rho}{\mathrm{d}t}=\frac{i}{\hbar}\left[\hat{H},\rho\right]+\tilde{R}\cdot\rho.
\end{equation}
The first part of Eq.~\eqref{eq:Liouville_I} describes the coherent evolution during the mw pulse, while the second part describes the recombination process.
Assuming a resonant mw pulse which selectively excites one of the two weakly coupled spins, the rotating frame Hamiltonian is given by
\begin{equation}
\label{eq:H}
\hat{H} = \frac{\omega_\mathrm{Rabi}}{2}
\begin{pmatrix}
0 & 1 & 0\\
1 & 0 & 0\\
0 & 0 & 0
\end{pmatrix}
,
\end{equation}
with the angular Rabi frequency $\omega_\mathrm{Rabi}$.
In Eq.~\eqref{eq:H}, we have taken into account that state $\left|3\right\rangle$ is not paramagnetic and therefore unaffected by the mw pulse. 

To simplify the discussion, we further neglect the coherences between states $\left|1\right\rangle$ and $\left|3\right\rangle$ and between $\left|2\right\rangle$ and $\left|3\right\rangle$, since the recombination process is incoherent.
Writing the remaining terms of $\rho$ as a column vector $\tilde{\rho}=(\rho_{11},\rho_{12},\rho_{21},\rho_{22},\rho_{33})^\mathrm{T}$, the recombination operator $\tilde{R}$ becomes
\begin{equation}
\tilde{R}=
\setlength{\extrarowheight}{2pt}
\setlength{\arraycolsep}{0pt}
\begin{pmatrix}
-\frac{1}{\tau_\mathrm{p}} & 0 & 0 & 0 & \frac{1}{2\tau_\mathrm{g}}\\
0 & -\frac{\tau_\mathrm{p}+\tau_\mathrm{ap}}{2\tau_\mathrm{p}\tau_\mathrm{ap}}-\frac{1}{T_\mathrm{d}} & 0 & 0 & 0\\
0 & 0 & -\frac{\tau_\mathrm{p}+\tau_\mathrm{ap}}{2\tau_\mathrm{p}\tau_\mathrm{ap}}-\frac{1}{T_\mathrm{d}} & 0 & 0\\
0 & 0 & 0 & -\frac{1}{\tau_\mathrm{ap}} & \frac{1}{2\tau_\mathrm{g}}\\
\frac{1}{\tau_\mathrm{p}} & 0 & 0 & \frac{1}{\tau_\mathrm{ap}} & -\frac{1}{\tau_\mathrm{g}}
\end{pmatrix}
,
\label{eq:R}
\end{equation}
with the recombination time of parallel spin pairs $\tau_\mathrm{p}$, the recombination time of antiparallel spin pairs $\tau_\mathrm{ap}$, and formation time constants of new spin pairs $\tau_\mathrm{g}$ with $1/\tau_\mathrm{g}=1/\tau_\mathrm{ec}+1/\tau_\mathrm{hc}$, where $\tau_\mathrm{ec}$ and $\tau_\mathrm{hc}$ denote the time constants of an electron and hole capture process, respectively, as defined in Fig.~\ref{fig:Figure4}(a) and Refs.~\cite{Dreher2012,Hoehne_Timeconstants_2013}. We additionaly included the dephasing time $T_\mathrm{d}$ to account for the experimentally observed dephasing of the coherent spin oscillations [Fig.~\ref{fig:Figure3}(d),(e)], which we attribute to inhomogeneities in the driving mw magnetic field.

The operator $\tilde{H}\cdot\tilde{\rho}=\frac{i}{\hbar}\left[\hat{H},\rho\right]$ describing the coherent evolution of $\tilde{\rho}$ then takes the form
\begin{equation}
\tilde{H}=i\frac{\omega_\mathrm{Rabi}}{2}
\begin{pmatrix}
0 & 1 & -1 & 0 & 0\\
-1 & 0 & 0 & 1 & 0\\
1 & 0 & 0 & -1 & 0\\
0 & -1 & 1 & 0 & 0\\
0 & 0 & 0 & 0 & 0
\end{pmatrix}
.
\label{eq:H1}
\end{equation}

We numerically solve Eq.~\eqref{eq:Liouville_I} by calculating
\begin{equation}
\tilde{\rho}(t) = \left\{ 
\begin{array}{l}
e^{(\tilde{H}+\tilde{R})t}\cdot\tilde{\rho}(0)~\mathrm{during~the~mw~pulse~and}\\
 e^{\tilde{R}t}\cdot\tilde{\rho}(0)~\mathrm{after~the~mw~pulse,}
\end{array}
\right.
\end{equation}
with $\tilde{\rho}(0)$=$(1,0,0,0,0)^\mathrm{T}$ assuming that the spin system is in the parallel state at the beginning of the mw pulse.
Finally, we calculate the relative change of the resistance $\Delta R(t)/R=\Delta n/n$ with $n$ denoting the electron and hole density in the conduction and valence band. The change of $n$ due to the spin-dependent recombination is given by $\Delta n=\left(\tau_\mathrm{l}/\tau_\mathrm{g}\right)\cdot n_\mathrm{sp}\cdot \Delta \tilde{\rho}_{33}$, with $\Delta \tilde{\rho}_{33}(t)=\tilde{\rho}_{33}(t)-\tilde{\rho}_{33}(0)$, the carrier lifetime $\tau_\mathrm{l}$ and the total density $n_\mathrm{sp}$ of $^{31}$P-P$_\mathrm{b0}$ spin pairs.
With $n=G\cdot \tau_\mathrm{l}$~\cite{Hoehne_Timeconstants_2013}, the relative change of resistance is given by
\begin{equation}
\frac{\Delta R(t)}{R} =\frac{n_\mathrm{sp}}{G\cdot \tau_\mathrm{g}}\Delta \tilde{\rho}_{33}(t),
\label{eq:DeltaR}
\end{equation}
where $G$ denotes the excitation rate of carriers by the above-bandgap illumination~\cite{Hoehne_Timeconstants_2013}.
The resulting $\Delta R(t)/R$ is plotted in Fig.~\ref{fig:Figure4}(a), 
using the parameters $\tau_\mathrm{p}$=1200~\textmu s, $\tau_\mathrm{ap}$=2~\textmu s, $G=5\cdot 10^{20}$cm$^{-3}$s$^{-1}$~\cite{Hoehne_Timeconstants_2013}, while $\tau_\mathrm{g}$=2.6~\textmu s, $T_\mathrm{d}$=210~ns and $n_\mathrm{sp}= 3\cdot 10^{12}$cm$^{-3}$ are used as fitting parameters to match the experimental data in Fig.~\ref{fig:Figure3}(a). 

The simulated transient reproduces the basic features of the experimental data in Fig.~\ref{fig:Figure3}(a) with characteristic rise and fall times determined mainly by $\tau_\mathrm{ap}$ and $\tau_\mathrm{g}$, respectively. Again, the coherent oscillations during the mw pulse are revealed after subtraction of a second order polynomial background as shown in Fig.~\ref{fig:Figure4}(c). The oscillation amplitude of $\sim 2\cdot 10^{-5}$ is a factor of $\sim$40 smaller compared to the simulated maximum total change of the resistance in good agreement with the experimentally observed suppression by a factor of $\sim$20. We therefore conclude that the time constants of the recombination process naturally explain the observed shape of the transient as well as the amplitude of the coherent oscillations. For a more detailed modeling, a distribution of recombination and generation time constants over the spin pair ensemble has to be taken into account~\cite{Hoehne_Timeconstants_2013}.

%The signal-to-noise ratio of pulsed rf-EDMR shown in Fig.~\ref{fig:Figure3}(b) is until now significantly lower compared to conventional pulsed EDMR measurements of the same sample in contrast to the improvement observed in cw rf-EDMR of Si field-effect transistors~\cite{Huebl09}. We tentatively attribute this to low-frequency fluctuations of the sample resistance caused by the mw pulses and the illumination~\cite{Hoehne2012}. Therefore, it would be interesting to apply pulsed rf-EDMR techniques to investigate spin-dependent transport processes in two-dimensional electron gases~\cite{Graeff1999,Lo2011}, where in addition to the increase in time resolution an improvement in the signal-to-noise level is expected as well~\cite{Huebl09}. 

In conclusion, we implemented rf-reflectometry readout for pulsed EDMR thereby increasing the measurement bandwidth by almost two orders of magnitude compared to current preamplifier-based detection schemes. This opens the way to studying faster charge dynamics, e.g., in direct semiconductors which with very few exceptions~\cite{Carlos1995,Bayerl1997,Wimbauer1998} so far have almost exclusively been studied by optically detected magnetic resonance because of their shorter carrier life times compared to indirect semiconductors such as silicon. Other systems that might benefit from the increased bandwidth are formation and dissociation processes of spin pairs in organic semiconductors~\cite{Grozema2002,Virgili2003} and donor-bound excitons in silicon~\cite{Schmid1977,Steger2012}. Furthermore, when applying rf-reflectometry to device structures like diodes~\cite{Carlos1995,Bayerl1997} or two-dimensional electron gases~\cite{Graeff1999,machida2003,Lo2011}, where no illumination is needed for EDMR measurements, a significant reduction of the noise level is expected since rf-reflectometry is less sensitive to low-frequency noise~\cite{Huebl09}. In particular, when spin-dependent scattering processes are detected, the large sensitivity of EDMR and the high time resolution demonstrated here might enable the observation and feedback control of spin fluctuations in small spin ensembles~\cite{Budakian2005,Goennenwein2000}.  

%with shorter carrier life times compared to silicon~\cite{Minsky2002}, where in particular recombination processes in nitride-based optoelectronic devices are studied by magnetic resonance methods~\cite{Carlos1995}.

 %Using this technique, we electrically detected in real time coherent spin oscillations in the test-bed of phosphorus-doped silicon. 
%The combination of spectroscopic information and nanosecond time resolution obtained by the presented pulsed rf-EDMR technique allows to study fast charge dynamics like, e.g., donor-bound excitons~\cite{Schmid1977} in silicon polaron formation and dissociation in organic semiconductors~\cite{Grozema2002,Virgili2003}. Further, it is applicable also to direct semiconductors 

%In addition to this, the combination of rf-reflectometry and pulsed EDMR can be useful to study fast recombination processes, e.g., in organic semiconductors~\cite{Grozema2002,Virgili2003} or direct semiconductors and is compatible with current device designs. 

The work was financially supported by DFG (Grant
No. SFB 631, C3).
% and the German Excellence Initiative via the "Nanosystems Initiative Munich" (NIM).   

%merlin.mbs apsrev4-1.bst 2010-07-25 4.21a (PWD, AO, DPC) hacked
%Control: key (0)
%Control: author (8) initials jnrlst
%Control: editor formatted (1) identically to author
%Control: production of article title (-1) disabled
%Control: page (0) single
%Control: year (1) truncated
%Control: production of eprint (0) enabled
%

\end{document}